\documentclass[ showpacs,preprint]{revtex4}
\usepackage[dvips]{epsfig,graphicx}
\usepackage{float}
\usepackage[dvipdfm,colorlinks=true, citecolor=blue, urlcolor=blue ]{hyperref}
\usepackage[mathscr]{euscript}
\usepackage{amsfonts}

\usepackage[mathscr]{euscript}
\usepackage{amsfonts}

\def\la{\langle}
\def\ra{\rangle}

\begin{document}
\title{Critical Casimir Interactions
and Percolation: the quantitative description of critical fluctuations.}

\author{Oleg A. Vasilyev}
\affiliation{Max-Planck-Institut f{\"u}r Intelligente Systeme,
  Heisenbergstra{\ss}e~3, D-70569 Stuttgart, Germany}
\affiliation{ IV. Institut f{\"u}r Theoretische Physik,
  Universit{\"a}t Stuttgart,  Pfaffenwaldring 57, D-70569 Stuttgart, Germany}

\begin{abstract}
Casimir forces in a critical media 
are produced by the spatial suppression 
of order parameter fluctuations.
In this paper we address the question how fluctuations of
a critical media relate the magnitude of  critical Casimir interactions.
Namely, for the Ising model we express the potential of 
 critical Casimir interactions  in terms of  Fortuin-Kasteleyn 
site-bond correlated percolation  clusters. 
These clusters are quantitative representation of fluctuations in a media.
New Monte Carlo method 
for the computation of the Casimir force potential which is  based
on this relation  is proposed.
We verify this method by computation of Casimir interactions
between two disks for 2D Ising model.
The new method is also applied to the investigation of 
non-additivity of the critical Casimir potential.
The non-additive contribution to three-particles interaction is
computed as a function of the temperature and the separation between disks. 
The benefit of the proposed method
is that it let us to compute the multi-particle interaction explicitly.
\end{abstract}

\pacs{05.70.Jk, 64.60.Ak, 05.10.Ln}

\maketitle
\section{Introduction}
The critical behavior in the vicinity of the second
order phase transition  
point is characterized by  large scale fluctuations
of the order parameter. The appearance of forces due
to the spatial confinement of fluctuations in the vicinity of 
the critical point
has been predicted by Fisher and de~Gennes~\cite{FdG}.
This phenomenon 
in a critical media  is now known as the {\it Critical Casimir} (CC) 
effect~\cite{Krech,BDT,Gambassi}. 
 One can observe the appearance of CC forces  in
 a critical binary mixture near its consolute point~\cite{Fukuto}. 
 Two colloidal particles in a fluctuating media 
exhibit attraction.
 The observation of colloidal particles 
aggregation in the critical binary mixture
has been described in~\cite{BE}.
The interaction  between 
a colloidal particle and a planar substrate 
has been measured directly my means of 
Total Internal Reflection Microscopy~\cite{nature,PRE1}.
Critical depletion in colloidal suspensions has been studied 
in many experiments~\cite{BCPP1,BCPP2,MKG,NFHW,DVN} 
(see Ref.~\cite{NDNS} for review).
Recently, the non-additivity of CC interactions between
three colloidal particles has been studied experimentally~\cite{NAE}.

The analytical computation of CC forces is a difficult problem.
A critical binary mixture belongs to the universality class of 
the Ising model. Therefore one can use numerical simulations
of the Ising model to extract information about CC interactions
for particular geometry and boundary conditions.
     The CC forces and its scaling functions for 3D Ising universality class
 with and without the bulk field
 for the film geometry and  various boundary conditions 
 have been studied numerically~\cite{EPL,PRE,hasen1,Surf,VD,CJH}.
     CC force between a spherical particle and a plane 
 for the 3D Ising universality class  has been studied in Ref.~\cite{Hsp}.
The numerical algorithm for the computation of CC interactions
between two particles in the presence of the 
negative bulk field for 3D Ising model has been proposed in~\cite{Vas}
and the interaction of a particle with two walls has been studied in~\cite{VM}.

Numerical methods also  are used for the computation of 
critical interactions within mean field (MF) universality class.
 The CC force between two colloidal particles 
in 3D has been first studied  
 using the conformal transformation in Refs.~\cite{ER,SHD}.
 MF interactions between an elliptic   
particle and a wall~\cite{KHD}
and between two colloidal particles
in the presence of the bulk ordering field~\cite{MKMD}
 have been studied. Non-additive interactions for MF universality class
 have been investigated for plane-particle-particle~\cite{MHD1}
 and three-particles~\cite{MHD2} geometries.

Experimentally has been shown,
that in a lipid membrane consisting of two different types
of lipids the second order phase transition may occur~\cite{VCSHHB,MPSV}. 
Such transition has been observed in a membrane forming
plasma vesicles~\cite{GKMV}. 
These experimental observations
demonstrate the  request for studying of the critical behavior
of the 2D Ising model to which universality class they belongs. 
For 2D Ising model for the stripe geometry the CC force may be 
computed analytically~\cite{DME1,DME2,AM1,AM2}.
       Results for  CC force between two disks 
 for the 2D Ising model  with the bulk field have been obtained via
 Derjuaguin approximation~\cite{ZMD}. 
 The alternative method 
 for the computation of the CC interaction 
   between a mobile disk and a wall has been recently proposed~\cite{HH}. 
The interaction between inclusions 
in a critical 2D  membrane has been studied in~\cite{MVS} by using the 
 Bennett method. The torque, acting on a needle near a wall
 has been studied in Ref.~\cite{VED}.
 The phase diagram of ternary solvent-solvent-colloid
mixture represented by 2D Ising model with disk-like
particles has been investigated by using
grand-canonical insertion technique~\cite{ETBERD}.
Recently, the fluctuation induced Casimir interactions
in colloidal suspensions at the critical point 
in 2D system has been studied
by geometric cluster algorithm~\cite{HH2}.
This algorithm capable to provide moving the particles 
and mixing the medium. Then two-particle
and multi-particle interactions are extracted from 
particle distribution functions. For 2D Ising model
at criticality one can use the power of conformal transformations
for analytical evaluation of CC interactions
between inclusions~\cite{GKMV,HH2}.
Recently, conformal invariance has been used for 
the investigation of interactions between rod-like particles~\cite{EB}.

In this article we propose the short and elegant expression 
for CC interactions between objects based on the
of critical percolation clusters.
The introduced method is based on counting of a number of percolation clusters
``touching'' immersed objects.
The paper is organized as follows: in the next section we 
express the interaction potential of CC force in terms
of clusters of Fortuin-Kasteleyn site-bond percolation.
We verify the numerical method by computing CC interactions
between two disks and comparing with results of another numerical approach. 
Then the proposed method
is applied to the computation of the non-additive interactions 
between three particles as a function of the inverse temperature. 
We and with a conclusions.
\section{Casimir interaction expressed via critical percolation clusters}
Let us consider the Ising model on a simple square  lattice
with periodic boundary conditions.
All distances are measured in lattice units.
The classical spin $\sigma_{i}= \pm 1$ is 
located in a site $i$ of the lattice. 
The inverse temperature is $\beta=1/(k_{\mathrm B}T)$.
The standard Hamiltonian of a bulk system (no restrictions for 
a spin directions)
for a spin configuration $\{ \sigma \}$ reads
\begin{equation}
\label{eq:H}
{\cal H}_{\mathrm b}(\{ \sigma \})=
-J \sum \limits_{\la ij \ra }\sigma_{i} \sigma_{j},
\end{equation}
where $J=1$ is the interaction constant, the sum $\la ij \ra$
is taken over all pairs of neighbor spins. The partition function of the 
model is given by the sum over the total set $\la \sigma \ra$
of all spin configurations 
$Z_{\mathrm b}=\sum_{\la \sigma \ra}
{\mathrm e}^{-\beta {\cal H}_{\mathrm b}(\{ \sigma \}) }$.
The corresponding free energy of the bulk system is 
$F_{\mathrm b}(\beta)=-\frac{1}{\beta} \ln[Z_{\mathrm b}(\beta)]$.
One can rewrite the partition function in terms of the 
Fortuin-Kasteleyn correlated site-bond percolation~\cite{FK}
(reviews on percolation theory are Refs.~\cite{Essam,SA,Saberi}).
For every bond between two spins $\sigma_i$
and $\sigma_j$ we introduce the bond variable $n_{ij}$,
which can take values 0 (open) and 1 (closed).
We introduce the probability $p=1-{\mathrm e}^{-2 \beta}$
for a bond between two parallel $\sigma_{i}=\sigma_{j}$ 
spins to be closed $n_{ij}=1$. In 
our case of homogeneous interactions with
the fixed constant $J=1$ the value of this probability $p$
does not depend on spin indexes $ij$.
The probability for 
a  bond between parallel spins to be open
$n_{ij}=0$ is $1-p={\mathrm e}^{-2 \beta}$
(the concept of open and closed bonds between parallel spins
has been proposed by Coniglio and Klein~\cite{CK}). 
A bond between two antiparallel $\sigma_{i}=-\sigma_{j}$ 
spins is always open $n_{ij}=0$. We denote $\{ n\}$ the 
particular configuration of bond variables. 
The standard expression for the partition function
may be rewritten  as
the sum over all set $\la n \ra $ of configurations of open
and closed bonds
\begin{equation}
\label{eq:Z2}
\begin{array}{c}
Z_{\mathrm b}(\beta)=\sum \limits_{\la  \sigma \ra}
{\mathrm e}^{-\beta {\cal H}_{\mathrm b}(\{ \sigma \}) }=
\sum \limits_{\la \sigma \ra} \prod \limits_{\la ij \ra }
{\mathrm e}^{\beta  \sigma_{i}\sigma_{j} }=\\
\rule{0pt}{15pt}  
=
{\mathrm e}^{\beta N_{b}}
\sum \limits_{\la \sigma \ra}
\sum \limits_{ \la n \ra} \prod \limits_{\la i j \ra}
\left[(1-p)\delta_{n_{ij},0}+
p\delta_{n_{ij},1}\delta_{\sigma_{i},\sigma_{j}} \right] =\\ 
\rule{0pt}{15pt} 
={\mathrm e^{\beta N_{b}}}
\sum \limits_{ \la n \ra} \left[ 
p^{n_{c}(\{n \})} (1-p)^{N_{b}-n_{c}(\{n \})}
2^{c(\{n \})}\right],
\end{array}
\end{equation}
where $N_{b}$ is the total number of bonds in the system,
$n_{c}(\{n \})$ is the number of closed bonds 
and $c(\{n \})$ is the number of cluster of connected spins
in the configuration $\{n \}$,
$2=q$ is the number of 
spin states for the  Ising model. The set $\la n \ra$ consists of $2^{N_b}$
different bond configurations.
\begin{figure}[h]
\begin{center}
\includegraphics[width=0.38\textwidth]{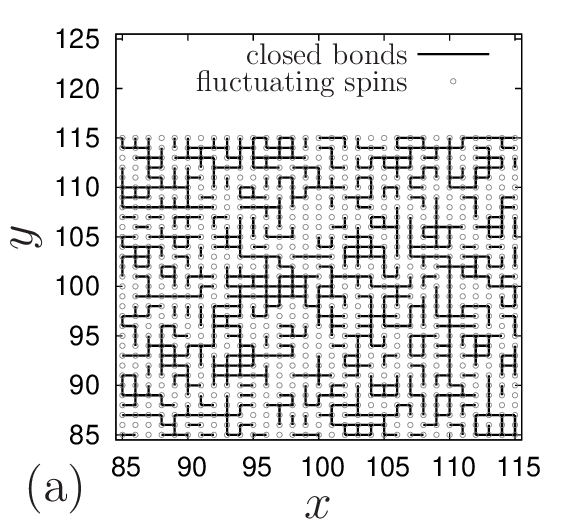}
\includegraphics[width=0.38\textwidth]{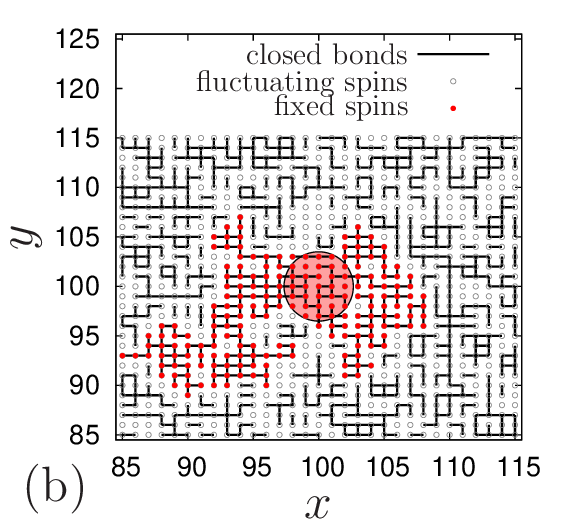}
\end{center}
\caption{%
(Color online)
(a) Clusters of spins, connected by closed bonds.
    Each cluster can take either $-1$ or $+1$ orientations, these fluctuating
    spins are denoted by empty circles.
(b) The same configuration of clusters with the inserted 
disk of the radius $R=2.5$ with the center at the point $(100,100)$.
All  clusters  which contain spins within the circle  
should be fixed $+1$. Fixed spins of  
these clusters are denoted by filled circles.
  }
\label{fig:col}
\end{figure}
Clusters of parallel spins,
connected by closed bonds are called ``physical'' 
clusters (contrary to ``geometrical'' clusters of parallel spins).
Detailed derivation of the partition function
and expressions of physical quantities
like energy, specific heat, magnetization, and 
magnetic succeptibility in terms of physical clusters
 have been provided by De~Meo, Heermann and Binder~\cite{OHB}.  
The Fortuin-Kasteleyn correlated site-bond percolation
is the basis of the  Swendsen-Wang numerical cluster algorithm
for spin models~\cite{SW}.
We write  eq.(\ref{eq:Z2}) in the form of the Ising model
(e.g., $\sigma_{i}= \pm 1$, $q=2$ is the number of spin states). 
It is also possible to rewrite
it for $q$-state Potts model ($\sigma_{i}=0,1,\dots,q$) 
with an arbitrary value of $q$. It means, that this approach 
generalizes  CC interactions for the Potts model.

Let us clarify the definition of a cluster of connected spins.
Two spins $\sigma_{i}$ and $\sigma_{j}$ are connected if the bond
between these spins is closed $n_{ij}=1$,
otherwise  these spins are not connected. 
The set of all spins, connected between each other
by closed bonds is called the cluster. 
In accordance with the definition of 
closed bonds, all spins in a given cluster are parallel.
But not every pair of parallel neighbor spins belong
to the same cluster. Let us note, that in the last line
of eq.~(\ref{eq:Z2}) the sum is taken only 
over the set $\la n \ra$ of all possible 
 configurations of open and closed bonds 
 (totally $2^{N_{b}}$ configurations). 
 At the same time for every particular configuration
 $\{ n \}$ the contribution from all possible 
 $-1$ and $+1$ cluster configurations is given by the term
  $q^{c(\{n \})}=2^{c(\{n \})}$ which counts the number
  of clusters $c(\{n \})$ for this configuration. The example of 
  such bond configuration on the square lattice
  is given in Fig.~\ref{fig:col}(a)
  where closed bonds are shown by solid lines
  and open bonds are not indicated.
 Directions of all spins in Fig~\ref{fig:col}(a) 
 are not specified because each cluster 
 can take both possible directions $-1$ and $+1$. 
Let us now consider the system with immersed
 particle (or particles). 
In terms of the Ising model it means, that 
all spins of such particle are 
fixed~\cite{Hsp,Vas,HH,HH2}. 
In Fig.~\ref{fig:col}(b) the example of the disk 
of the radius $R=2.5$ with the center 
located in the point $(100,100)$ is demonstrated.
We denote $\{{\mathrm{col}}\}$ the set of all fixed spins 
in our colloidal particle(s), in the current example
$\sigma_{k}=+1$ for spins with coordinates 
$(x_{k}-100)^{2}+(y_{k}-100)^{2} \le R^2$.
The partition function $Z_{c}$ of the system with immersed 
colloidal particle(s) may be expressed via the Hamiltonian
of a bulk system ${\mathcal H}_{\mathrm b}$ with the application
 of the constraint $\delta_{\sigma_{k},1}$ to all fixed spins.
 On the other hand, this partition function may be expressed 
 via the third line of eq.~(\ref{eq:Z2}) if we take into account,
 that all clusters $c_{c}$ which contain spins of the colloidal particle
 can not fluctuate. The partition function of 
 the system with immersed particle(s) 
 is expressed as
\begin{equation}
\begin{array}{c}
Z_{c}(\beta)=
\sum \limits_{\la \sigma \ra }\prod 
\limits_{ k  \in \{ {\rm col} \} }
\delta_{\sigma_{k},1}
  {\mathrm e }^{-\beta {\cal H}_{\mathrm b}(\{ \sigma  \}) }=\\
\rule{0pt}{16pt}
={\mathrm e^{\beta N_{b}}}
\sum \limits_{ \la n \ra}\left[ p^{n_{c}
(\{n \})}
(1-p)^{N_{b}-n_{c}(\{n \})}2^{c(\{n \})-c_{c}(\{n \})}\right].
\end{array}
\end{equation}
The only difference from the partition function
of   the bulk system without particle(s) $Z_{\mathrm b}$
in eq.~(\ref{eq:Z2}) is the term $2^{-c_{c}(\{n \})}$.
This  term reflects the fact, that all $c_{c}$ clusters of the current 
configuration $\{n \}$, which contain spins of colloidal particle, 
are fixed and do not contribute to the partition function.
The free energy of a system with a particle(s) 
is $F_{c}(\beta)=-\frac{1}{\beta} \ln[Z_{c}(\beta)]$.
Therefore, we can expressed the free energy difference 
(in $k_{\mathrm B}T $ units) of the insertion of particle(s) as
\begin{equation}
\label{eq:dF}
\beta [F_{c}(\beta)-F_{\mathrm b}(\beta)]
=-\ln \left[\frac{Z_{c}(\beta)}{Z_{\mathrm b}(\beta)}\right]=-\ln \la  2^{-c_{c}}\ra_{\beta},
\end{equation}
where $\la  2^{-c_{c}}\ra_{\beta}$ is  2 
to the power $-c_{c}$ where $c_{c}$ is the number of clusters, touching
the immersed object 
computed as a thermal average $\la \dots\ra_{\beta}$
 at the inverse temperature $\beta$
with respect to the bulk (empty) system.
The proposed method can also be applied to 
systems with various types of boundary conditions and the presence of the bulk field.
 Let us note, that we do not specify how many particles
 are immersed into a bulk system.
 It means, that the simple  expression 
 eq.~(\ref{eq:dF}) may be applied to computation of CC
 interactions for various geometries: particle-particle,
 multi-particles, wall-particle,  etc. Moreover, this numerical method
 can be also applied to an arbitrary spatial dimension 
 and for $q \ne 2$ Potts model. 

\section{Model description and numerical verification}
Let us demonstrate the application of the proposed method to
the computation of CC interactions between two disks in 2D Ising system.
This example has a practical interest, because
it describes the interaction
between protein inclusions in a lipid membrane
at critical concentration of membrane 
components~\cite{VCSHHB,MPSV,GKMV,HH2,ETBERD,ZMD}.

\begin{figure}[!ht]
\begin{center}
\includegraphics[width=0.38\textwidth]{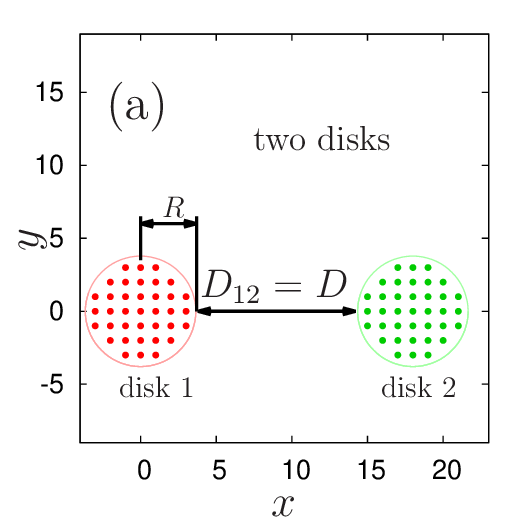}
\includegraphics[width=0.38\textwidth]{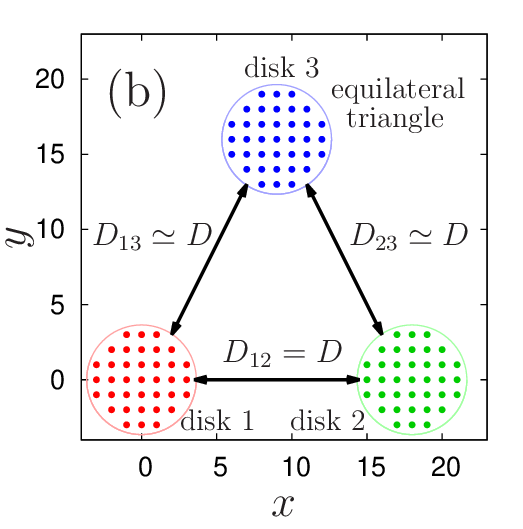}
\end{center}
\caption{%
(Color online) The system geometry for:
(a)  Two disks of the radius $R=3.5$ at the separation $D_{12}=D$
for the numerical verification of the algorithm;
(b) Equilateral triangle of disks
at separations $D_{12}=D$, $D_{13}=D_{23} \simeq D$
for the studying of non-additive interactions.
  }
\label{fig:geom}
\end{figure}

We consider the 2D  Ising model on a square lattice
with the periodic Boundary Conditions (BC).
All distances are measured in lattice units,
the system size is $200\times 200$.
For the reference bulk system with the free energy 
$F_{\mathrm b}(\beta)$ in
each site of the lattice the fluctuating spin 
$\sigma_{i}= \pm 1$ is located. For the system 
with immersed particles with the free energy $F_{c}(\beta,D)$
two disks (disk 1 and disk 2) of the half integer radius $R=3.5$
are immersed in the system at the distance $D$ -- see Fig.~\ref{fig:geom}(a).
Spins in these disks are fixed to be $+1$. 
This choice corresponds to the symmetry-breaking 
BC with completely ordered surface 
and it is usually denoted as $(++)$ BC~(see~\cite{Surf} for details). 
Let us denote $c_{12}(D)$ the number of all clusters, which 
contain spins of two disks at the distance $D$.
We denote the insertion free energy difference for two disks
\begin{equation}
 U_{\mathrm{ins}}^{12}(\beta,D)=
 \beta[F_{c}(\beta,D)-F_{\mathrm b}(\beta)]=- 
 \ln \la 2^{-c_{12}(D)}\ra_{\beta},
\end{equation}
which depends on the distance between disks $D$
and is expressed via the logarithm of the average value of $2$
to the power of $-c_{12}(D)$ in accordance with eq.(\ref{eq:dF}).
The insertion energy itself does not provide the information
about disks interaction because it is defined up to a certain constant.
Let us take as a normalization constant the insertion energy
$U_{\mathrm{ins}}^{12}(\beta,D_{\mathrm{max}})$ which corresponds to the maximal possible
separation $D_{\mathrm{max}}$ between particles for the finite system
with periodic BC. In our case $D_{\mathrm{max}}=L-2R$
where $L=200$ is the system size.
We introduce the CC interaction potential
\begin{equation}
\label{eq:Ucas}
\begin{array}{c}
U_{12}(\beta,D)= U_{\mathrm{ins}}^{12}(\beta,D)-
U_{\mathrm{ins}}^{12}(\beta,D_{\mathrm{max}})=\\
\rule{0pt}{16pt}
=\ln \la 2^{-c_{12}(D_{\mathrm{max}})}\ra_{\beta}- \ln \la 2^{-c_{12}(D)}\ra_{\beta},
\end{array}
\end{equation}
which derivative with respect to the distance provides the Casimir
force $f_{\mathrm{Cas}}=-\partial U_{12}/\partial D$
between two disks.

\begin{figure}[h]
\begin{center}
\includegraphics[width=0.38\textwidth]{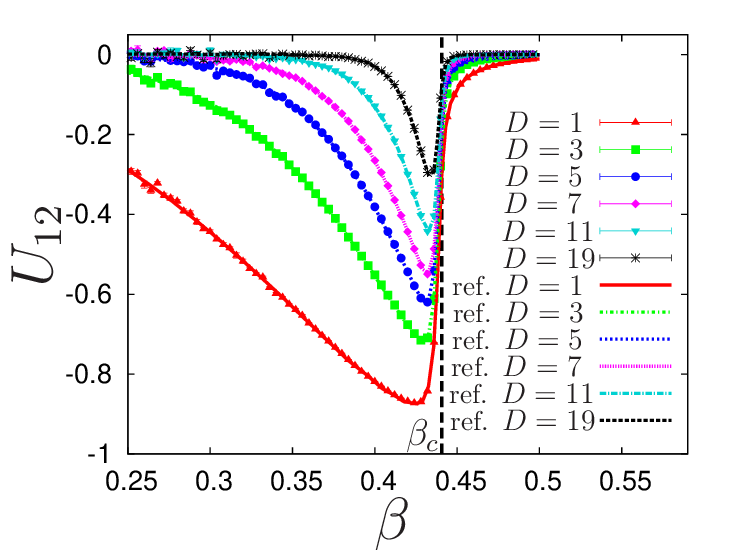}
\end{center}
\caption{%
(Color online)
 Numerical verification of CC interaction potential $U_{12}$
as a function of the inverse temperature $\beta$ 
between two disks at distances $D=1,3,5,7,11,19$ computed 
via percolation clusters (symbols) and 
by using the reference method~\cite{Vas}~(lines). 
  }
\label{fig:check}
\end{figure}

 As a first step we numerically verify the algorithm
by computing the interaction potential $U_{12}(\beta,D)$
between disk 1 and disk 2 in accordance with eq.~(\ref{eq:Ucas}).
We use the hybrid Monte Carlo (MC) algorithm~\cite{LB}.
Each step consists of one update in accordance with
 Swendsen-Wang algorithm~\cite{SW} (which is efficient
in the vicinity of the critical point) followed by $L^2/5$
attempts of Metropolis spin updates~\cite{Met} at random positions
(which is efficient out of criticality).
 The averaging is performed over $8 \times 10^{8}$ MC steps
splitted on 10 series for the evaluation of the numerical inaccuracy.
Let us note, that in accordance with our
algorithm we can simultaneously perform computations
for a set of disks separations $D$.
In Fig.~\ref{fig:check} we plot by symbols the interaction potential
$U_{12}$ between two disks as a function of the inverse temperature $\beta$
for various separations $D=1,3,5,7,11,19$.
In the same figure we also plot by lines 
the potential computed by alternative reference method.
The reference method (ref.) is based on 
the numerical integration of local magnetization over
the locally applied field~\cite{Vas,rev}.
We observe the perfect agreement between results computed
in two different ways. It confirms  eq.~(\ref{eq:Ucas})
and verifies the numerical code which is used for simulations. 
The position of the critical value of the inverse temperature 
$\beta_c=\ln(\sqrt{2}+1)/2$
is denoted by the vertical dashed line. The maximum of the attractive 
interaction is slightly shifted to high-temperature region
as it is typical for $(++)$ BC~\cite{Surf}. The proposed method 
can be straightforwardly expanded for systems with bulk and surface 
fields and systems with several values of the interaction constants.

\section{ Non-additivity of three particle interactions}

Let us now study the non-additivity of interaction in three-particles
system with the geometry of an equilateral triangle.
This geometry is shown in Fig.~\ref{fig:geom}(b) 
and it has been  used in experiment~\cite{NAE}
and in the investigation of non-additivity for MF universality class~\cite{MHD2}.
Now we add the third disk 3 at the separation
$D_{13}$ from the first disk and at the 
separation $D_{23}$ from the second disk.
The separations selected to be almost equal on the lattice
$D_{13}=D_{23} \simeq D_{12} \equiv D$ 
Later on for the equilateral triangle we select 
the odd distance $D$ so the projection 
of the center of the third disk on  $x$-axis
is located exactly between centers of disk 1 and disk 2.
\begin{figure*}[!ht]
\begin{center}
\mbox{\includegraphics[width=0.45\textwidth]{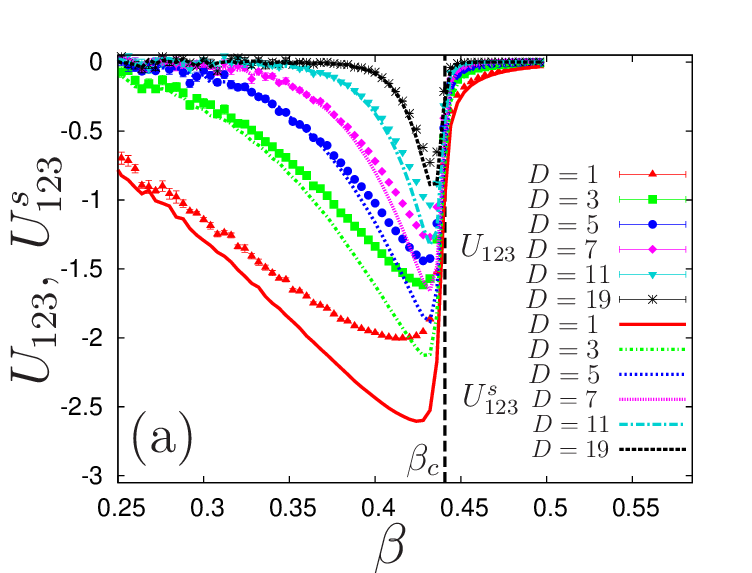}
\includegraphics[width=0.45\textwidth]{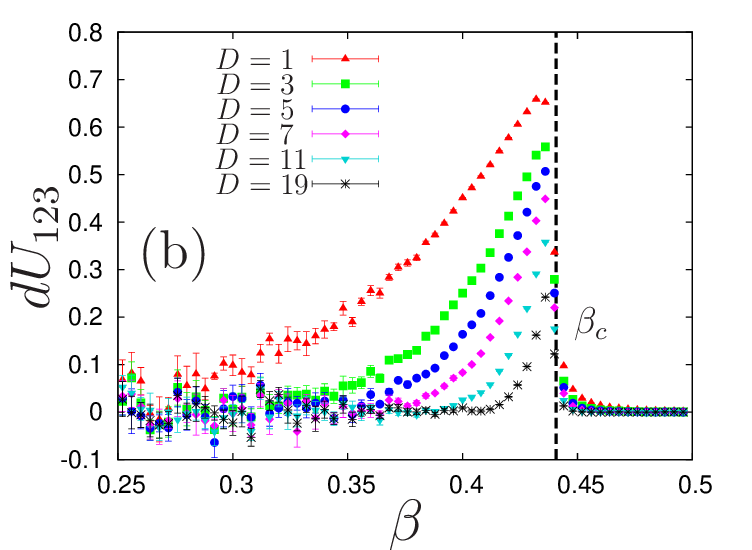}}
\end{center}
\caption{
(Color online) 
(a) Multiparticle CC interaction potential $U_{123}$ (symbols)
between three disks located in vertices of the
equilateral triangle with side of length $D=1,3,5,7,11,19$
as a function of the inverse temperature $\beta$ in comparison
with the sum of three pair potentials
$U_{123}^{s}=U_{12}+U_{13}+U_{23}$ (lines);
(b) Excess non-additive contribution to
CC interaction potential $dU_{123}=U_{123}-U_{123}^{s}$
as a function of the inverse temperature $\beta$
for three disks at same separations.
  }
\label{fig:res1}
\end{figure*}

\begin{figure*}[!ht]
\begin{center}
\mbox{\includegraphics[width=0.45\textwidth]{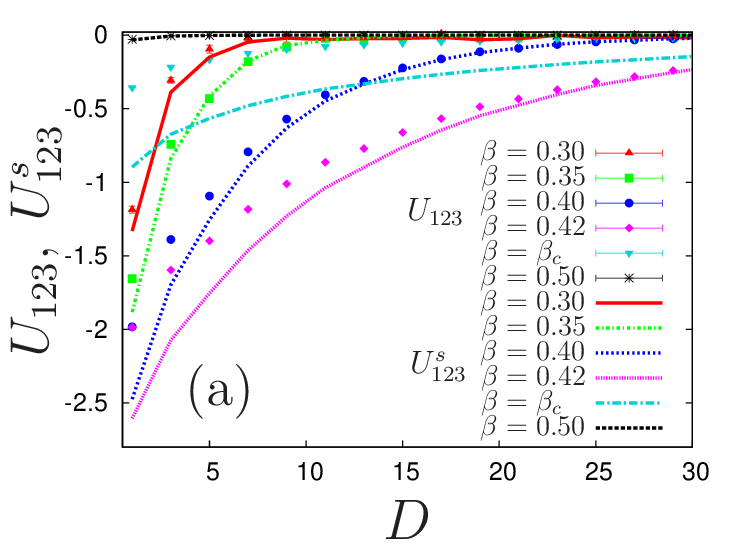}
\includegraphics[width=0.45\textwidth]{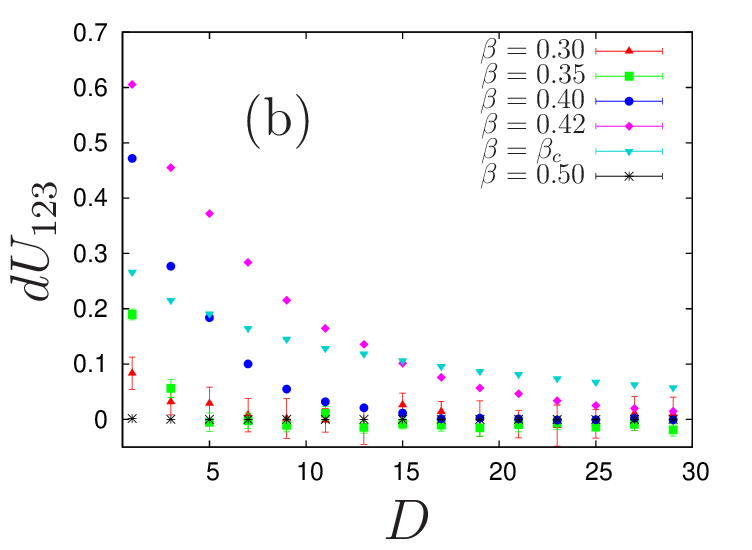}}
\end{center}
\caption{%
(Color online) 
(a) Multiparticle CC interaction potential $U_{123}$ (symbols)
between three disks located in vertices of the
equilateral triangle as a function of the side length $D$
for various values of the inverse temperature 
$\beta=0.3,0.35,0.4,0.42,\beta_{c},0.5$ in comparison
with the sum of three pair potentials 
$U_{123}^{s}=U_{12}+U_{13}+U_{23}$ (lines); 
(b) Excess non-additive contribution to 
CC interaction potential $dU_{123}=U_{123}-U_{123}^{s}$
for $\beta=0.3,0.35,0.4,0.42,\beta_{c},0.5$.
  }
\label{fig:res2}
\end{figure*}
We denote $c_{12}$, $c_{13}$, $c_{23}$, and $c_{123}$
the number of percolation clusters, which 
contain spins of disks 1 and 2, disks 1 and 3, disks 2 and 3,
and all disks 1,2,3, respectively.
The potential of three particle interaction 
is expressed as 
\begin{equation}
\label{eq:Ucas123}
\begin{array}{c}
U_{123}(\beta,D)
=\ln \la 2^{-c_{123}(D_{\mathrm{max}})}\ra_{\beta}- 
\ln \la 2^{-c_{123}(D)}\ra_{\beta}.
\end{array}
\end{equation}
We also introduce the sum of pair interactions 
$U^{s}_{123}=U_{12}+U_{13}+U_{23}$. For the additive potential we expect 
$U_{123}=U^{s}_{123}$.
In Fig.~\ref{fig:res1}(a) we plot $U_{123}$ as a function
of the inverse temperature $\beta$ for different separations
between disks $D=1,3,5,7,11,19$ by symbols.
In the same figure we plot the sum of pair potentials
$U^{s}_{123}$ by lines. The difference between $U_{123}$
and $U^{s}_{123}$ demonstrates us the non-additive nature 
of multiparticle CC interactions. Let us underline,
that for the fixed value of $\beta$ 
all quantities: $\la 2^{-c_{123}(D)}\ra_{\beta}$, $\la 2^{-c_{12}(D)}\ra_{\beta}$,
$\la 2^{-c_{13}(D)}\ra_{\beta}$, $\la 2^{-c_{23}(D)}\ra_{\beta}$
for all set of separations $D$ are computed during the single 
simulation of the bulk system! It means, that we should not perform 
separate simulation for each value of $D$.
In Fig.~\ref{fig:res1}(b) 
we plot the non-additive difference $dU_{123}= U_{123}-U_{123}^{s}$
as a function of $\beta$ for the same systems.
The non-additive part of interactions is positive.
It means, that the presence of the third particle
decreases the interaction between pair of particles. 
In Fig.~\ref{fig:res1}(b) we see, that the non-additive part of 
three-particle interactions has the maximum 
approximately at the same point, as the minimum of two-particle
interaction potential.
 All these potentials
have wide tails to the high-temperature region,
while interactions in low-temperature region 
decreases very fast as we separates from the critical point.
This behavior qualitatively coincide with
results for MF system~\cite{MHD1,MHD2} 
and with experimental results~\cite{NAE} for similar geometries.

In Figs.~\ref{fig:res2}(a)~and~(b) we plot multiparticle interaction potential
and its excess non-additive contribution
as functions of the separation $D$ for various values of the inverse temperature
$\beta$, respectively. Such presentation is more convenient
for experimentators, who typically measure the potential as a 
function of separation for the given temperature~\cite{nature}. 
The non-additivity of three-particle interaction
for 2D system at the critical point
has been previously studied in Ref.~\cite{HH2}.
In this article the information about interaction 
potentials is extracted from particle distribution functions.  
Contrary to this approach, our method provides the possibility to 
compute the interaction potential directly, simultaneously 
for several values of separations $D$.  

\section{Conclusion}
In the present paper we
express the free energy change for the insertion of objects
into a critical system in terms of percolation clusters 
intersecting the volume of inserting objects.
The numerical algorithm for CC interactions
which is based on counting of the number of percolation clusters
is proposed. The algorithm provides explicit expression 
for the Casimir potential without numerical integration.
This algorithm is numerically verified for 
the computation of CC force potential between two disks for 2D Ising model.
 The proposed method is also applied 
for the studying of multiparticle interactions
in a critical media and express the non-additive part of 
the interaction potential in terms of percolation clusters.
 Obtained expression
gives us the qualitative information 
about the sign of the non-additive contribution to multiparticle
interaction while the numerical realization of the proposed 
algorithm let us to perform quantitative computation of the interaction potential.
Results for the temperature dependence 
of the non-additive part of three-particle interaction for 2D system are 
presented first time.

\end{document}